\documentclass[
    ,final            % use final for the camera ready runs
  ]
  {aipproc}

\layoutstyle{8x11double}

\begin{document}

\title{An X-ray View of Radio Millisecond Pulsars}


\author{Slavko Bogdanov}{
  address={Harvard-Smithsonian Center for Astrophysics, 60 Garden Street, Cambridge, MA 02138}
}

\author{Jonathan E. Grindlay}{
  address={Harvard-Smithsonian Center for Astrophysics, 60 Garden Street, Cambridge, MA 02138}
}

%\author{Maureen van den Berg}{
%  address={Harvard-Smithsonian Center for Astrophysics, 60 Garden Street, Cambridge, MA 02138} % additional visiting address
%}

%\author{Craig O. Heinke}{
%  address={Harvard-Smithsonian Center for Astrophysics, 60 Garden Street, Cambridge, MA 02138}
%}

%\author{George B. Rybicki}{
%  address={Harvard-Smithsonian Center for Astrophysics, 60 Garden Street, Cambridge, MA 02138}
%}

%\author{Fernando Camilo}{
%  address={Columbia Astrophysics Laboratory, New York, NY}
%}

%\author{Paulo Freire}{
%  address={Arecibo Observatory}
%}

\begin{abstract}

  In recent years, X-ray observations with \textit{Chandra} and
  \textit{XMM-Newton} have significantly increased our understanding
  of rotation-powered (radio) millisecond pulsars (MSPs).  Deep
  \textit{Chandra} studies of several globular clusters have detected
  X-ray counterparts to a host of MSPs, including 19 in 47 Tuc
  alone. These surveys have revealed that most MSPs exhibit thermal
  emission from their heated magnetic polar caps.  Realistic models of
  this thermal X-ray emission have provided important insight into the
  basic physics of pulsars and neutron stars. In addition, intrabinary
  shock X-ray radiation observed in ``black-widow'' and peculiar
  globular cluster ``exchanged'' binary MSPs give interesting insight
  into MSP winds and relativistic shock.  Thus, the X-ray band
  contains valuable information regarding the basic properties of MSPs
  that are not accesible by radio timing observations.

\end{abstract}

\maketitle

\section{Introduction}

Rotation-powered (``recycled'') millisecond pulsars (MSPs) were
discovered at X-ray energies during the \textit{ROSAT} all-sky survey
\citep{Beck93}. In its lifetime, \textit{ROSAT} detected X-ray
emission from 10 MSPs in the field of the Galaxy and one in the
globular cluster M28 \citep{Beck99}. These observations established
that MSPs are very faint X-ray sources with typical X-ray luminosities
in the range $L_{X}\sim10^{30-31}$ ergs s$^{-1}$. In recent years, the
superb sensitivity of the
\textit{Chandra}\footnote{\url{http://cxc.harvard.edu/}} and
\textit{XMM-Newton}\footnote{\url{http://xmm.vilspa.esa.es/}} X-ray
observatories have permitted much more detailed studies of these
objects. \textit{Chandra}, in particular, with its unprecedented
sub-arcsecond angular resolution, has proven to be ideally suited for
studies of globular clusters, where MSPs are found in great
abundance. In addition, both facilities have allowed detailed
spectroscopic and timing studies of the nearby MSPs in the field of
the Galaxy \citep{Zavlin06,Zavlin07}.  Herein, we summarize the
current state of affairs and examine future prospects in the X-ray
study of MSPs.

\section{X-ray Studies of Globular Cluster Millisecond Pulsars}

Radio timing surveys have revealed that globular clusters are
veritable treasure troves of MSPs
\citep[e.g.,][]{Camilo00,Ransom05}. As such, they allow a very
efficient way to study MSPs at X-ray energies due to their inherent
faintness of these objects, which necessitates very long exposures.
This is best examplified by the detection of X-ray counterparts to all
19 MSPs with known radio timing positions (Fig. 1) in a deep
\textit{Chandra} ACIS-S exposure of 47 Tuc.  The large majority of the
47 Tuc MSPs (16 out of 19) appear to be soft, thermal X-ray sources
with temperatures $T_{\rm eff}\sim(1-3)\times10^6$ K, emission radii
$R_{\rm eff}\sim0.1-3$ km, and luminosities $L_{X}\sim10^{30-31}$ ergs
s$^{-1}$ \citep{Bog06a}.  The inferred effective emission areas imply
that this radiation originates from the heated magnetic polar caps of
the pulsar, as suggested by theoretical pulsar models
\citep{Hard02,Zhang03}. The thermal nature of the observed X-ray
emission was subsequently confirmed with \textit{Chandra} HRC-S timing
observations, which revealed relatively broad pulsations with
relatively low ($\le$50\%) pulsed fractions \citep{Cam07}.  For
several MSPs, there is indication of an additional thermal component
(implying non-uniform heating of the polar cap), similar to what is
seen in several nearby field MSPs \citep{Zavlin06}.  These
similarities in X-ray properties imply that, in general, no systematic
differences exist between MSPs in globular clusters and those in the
field of the Galaxy. It has been speculated that globular cluster MSPs
should differ from those in the field of the Galaxy due to repeated
dynamical binary exchanges and accretion episodes, which may alter the
magnetic field topology of the pulsar from a dipolar to a multipolar
one \citep{Grind02,Cheng03}. However, the X-ray data indicates that
even if most globular cluster MSPs do undergo ``re-recycling'', this
does not significantly alter the observable properties of these
objects.

\begin{figure}[t!]
  \includegraphics[height=.385\textheight]{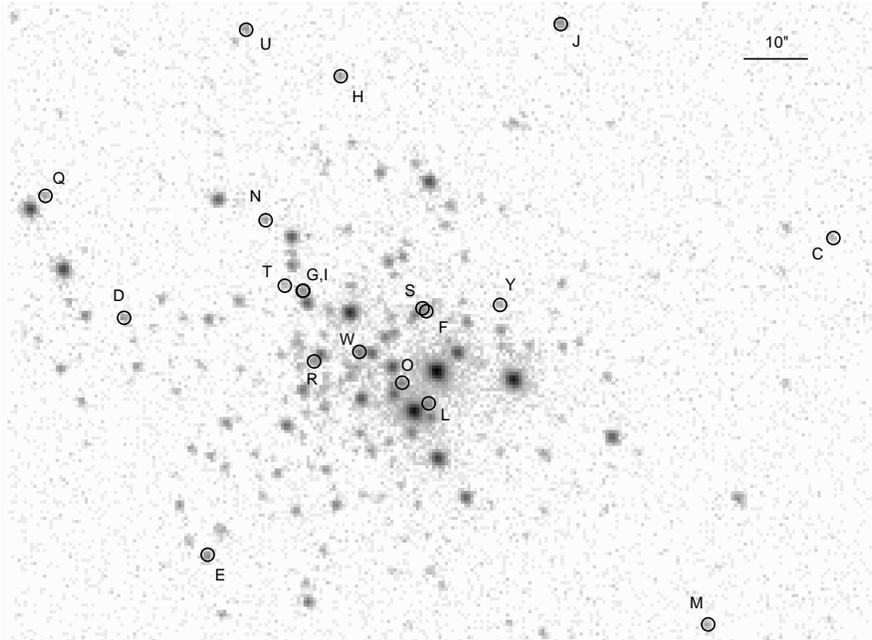}
  \caption{\textit{Chandra} ACIS-S 0.3--6 keV image of the core of the
  globular cluster 47 Tucanae. The 1'' circles are centered on the
  radio timing positions of the 19 MSPs. An X-ray source is associated
  with the position of each MSP \citep[see][for details]{Bog06a}.}
\end{figure}

\begin{figure}[t!]
  \includegraphics[height=.41\textheight]{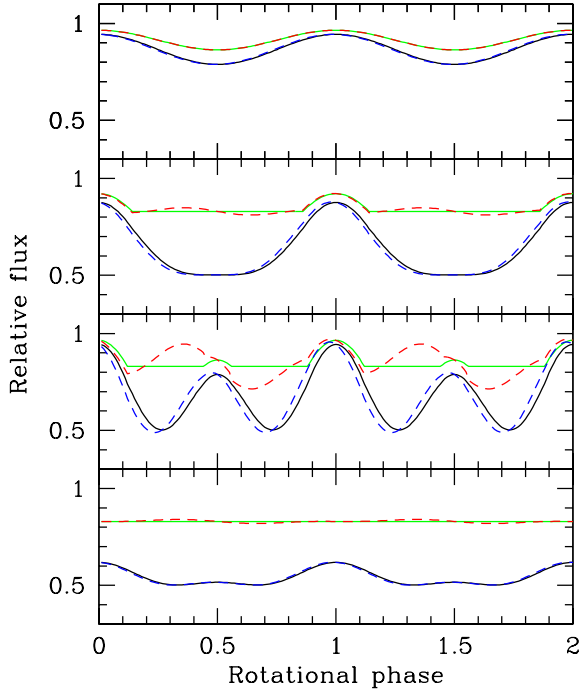}
  \caption{Representative hydrogen atmosphere model lightcurves for a
  rotating $M=1.4$ M$_\odot$, $R=10$ km NS with two point-like
  antipodal hot spots \citep{Bog07a}. The panels show pulse profiles for
  ($\alpha=10^{\circ}$, $\zeta=30^{\circ}$), ($\alpha=30^{\circ}$,
  $\zeta=60^{\circ}$), ($\alpha=60^{\circ}$, $\zeta=80^{\circ}$), and
  ($\alpha=20^{\circ}$, $\zeta=80^{\circ}$), from top to bottom,
  respectively. The solid lines in each plot correspond to a H
  atmosphere (\textit{black}) and blackbody emission (\textit{green})
  with no Doppler effect included. The dashed lines show the effect of
  Doppler boosting and aberration for $P=4$ ms.  All fluxes are
  normalized to the value for $\alpha=\zeta=0$.}
\end{figure}

\section{Modeling Thermal X-ray Emission from Millisecond Pulsars}

It has been shown that modeling X-ray spectroscopic and timing data of
thermal MSPs can be used to extract important information regarding
the properties of the neutron star \citep{Pavlov97,Zavlin98,Bog06b}.
Given that the vast majority of MSPs appear to be thermal X-ray
sources, it is important to develop realistic theoretical models to
ascertain whether this method can be employed to gain useful insight
into the properties of MSPs.  Such models of surface emission from
MSPs need to incorporate general relativistic effects of gravitational
redshift and bending of photon trajectories as well as special
relativistic effects (Doppler boosting and aberration) due to the
rapid motion of the stellar surface. It is also necessary to consider
an optically-thick, weakly magnetized light-element atmosphere
\citep[see, e.g.,][]{Zavlin96}. Such an atmospheric layer is expected at
the surface of a MSP given the standard formation scenaro of MSPs,
involving accretion of a substantial amount of gas \citep{Bhatt91}.

The synthetic pulse profiles of MSPs generated using these models are
characterized by broad pulsations and moderate pulsed fractions, which
are significantly larger than those of a blackbody emission spectrum
for the same assumed parameters (see Fig. 2).  Most notably, unlike
the beamed radio emission, the duty cycle of the X-ray radiation is
always 100\% and is observable for all combinations of $\alpha$ and
$\zeta$. These characteristics have important implications for
population studies of MSPs (see below).

\begin{figure}[t!]
  \includegraphics[angle=270,width=.38\textheight]{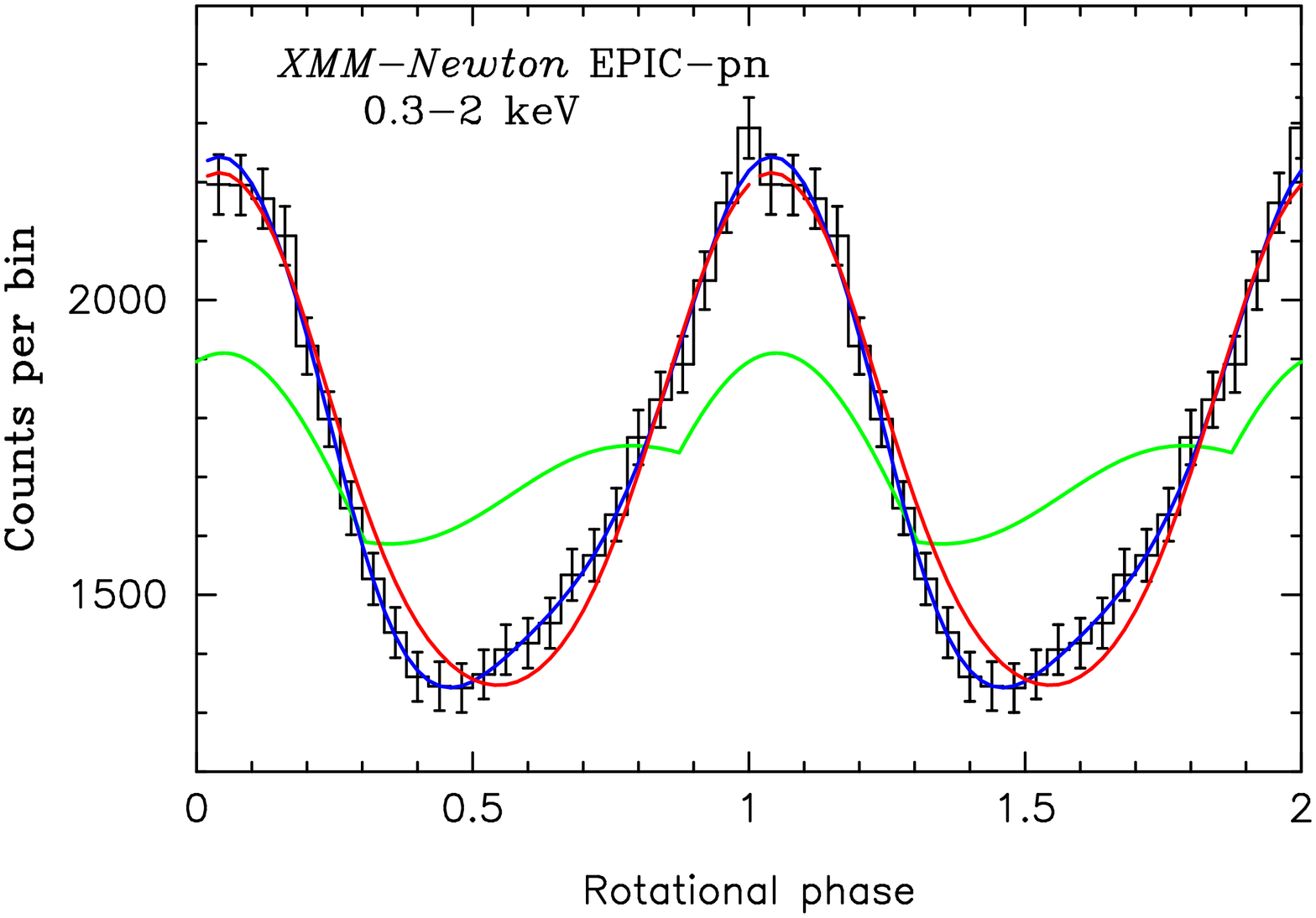}
\includegraphics[angle=270,width=.35\textheight]{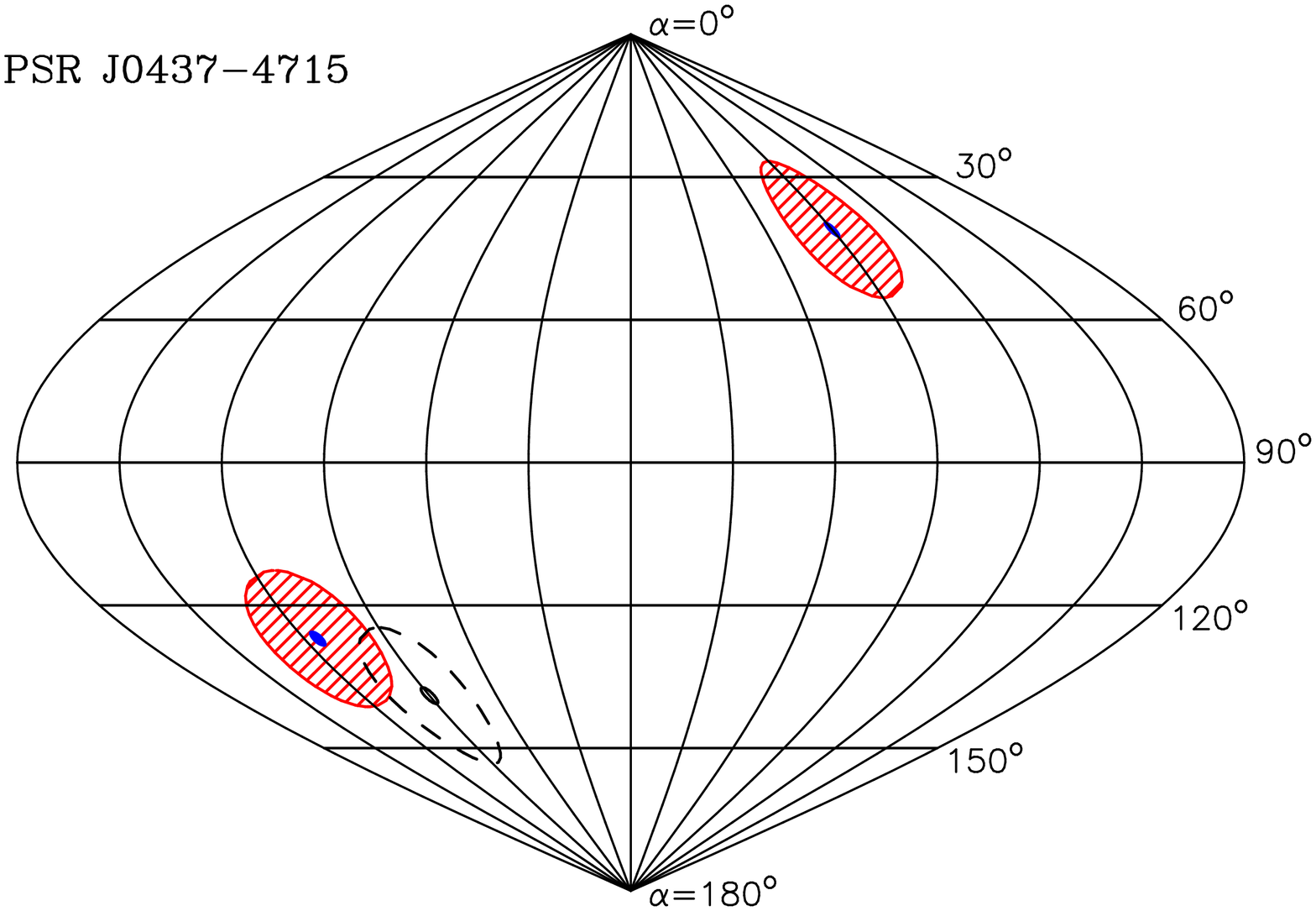}
\caption{(\textit{Left}) \textit{XMM-Newton} EPIC-pn 0.3--2 keV pulse
profile of PSR J0437--4715. The blue line shows the best fit curve for
an off-center dipole field and a H atmosphere, while the green line is
of a blackbody that best fits the spectrum of J0437--4715 for the same
assumed compactness and geometry. The red line corresponds to the best
fit H atmosphere model for a centered dipole (\textit{Right})
Equal-area sinusoidal projection surface emission map of PSR
J0437-4715 as inferred from the X-ray data for $R=10$ km and $M=1.4$
M$_{\odot}$. The dashed line shows the expected antipodal position of
the secondary polar cap.}
\end{figure}

A hydrogen atmosphere polar cap X-ray emission model has been found to
be in fair agreement with the available X-ray data of the three
nearest known MSPs, PSRs J0437--4715 \citep{Zavlin98,Bog07a},
J2124--3358, and J0030+0451 \citep{Bog07b}.  On the other hand, a
blackbody emission model is inconsistent the observed pulse profiles.
This provides compelling evidence that a gaseous atmosphere is indeed
present at the surface of these neutron stars. For PSRs J0437--4715
and J2124--3358, the data indicates that the two polar caps are not
diametrically opposite on the stellar surface, implying an off-center
dipole magnetic axis.  Perhaps more importantly, the thermal X-ray
pulsations offers interesting constraints on the mass-to-radius ratio
($M/R$) of the neutron star. In particular, for J0437--4715,
J2124--3358, and J0030+0451 we find $R=6.9-10.6$, $R>7.2$, and $R>9.5$
km, respectively (68\% confidence), assuming $M=1.4$ M$_{\odot}$.

Modeling of the thermal X-ray pulse profiles from MSPs also gives
useful limits on the magnetic field geometry of the pulsar, which in
turn provides stringest observational constraints on theoretical
pulsar models. For example, Ruderman has postulated the
presence of crustal ``plates'' on the NS surface formed by shear
stresses on the crust caused by neutron superfluid vortex lines pinned
to lattice nuclei. The motion of these plates would cause the magnetic
fields of MSPs to migrate across the stellar surface, resulting in
either an aligned magnetic field or one pinched at one of the spin
poles.  However, the observed thermal pulse shapes and pulsed
fractions from the nearest MSPs ($\sim$30--50\%) suggest that the
magnetic field resembles a (nearly) centered dipole and is moderately
inclined with respect to the spin axis (see Fig. 3).  This, in
turn, implies that the dipole fields of these (and likely all) MSPs do
not have a tendency to migrate towards the spin pole or align with the
spin axis.  Moreover, the infered X-ray--emitting areas are consistent
with the classical polar cap area [$R_{pc}=(2\pi R/cP)^{1/2}R$ km] and
do not favor so-called ``squeezed'' polar caps \citep{Rud91}.

%The fits to the
%thermal X-ray pulse profiles of MSPs offer useful insight into the
%magnetic field geometry of the pulsar. This has important implications
%for magnetic field evolution models. For instance, Ruderman (1991; see
%also Chen \& Ruderman 1993 and Chen et al. 1998) has postulated that
%the magnetic fields of MSPs migrate across the stellar surface during
%the spin-up (X-ray binary) phase so that the magnetic axis closely
%aligns with the spin axis or is squeezed to one of the spin
%poles. However, this configuration would result in very little (few
%percent) modulation of the thermal X-ray flux due to the close
%alignment of the polar caps with the spin axis, regardless of the
%viewing geometry. This is at odds with the observed pulsed fractions
%of PSRs J0437--4715, J0030+0451, and J2124--3358, which are in the
%range 30--50\%. Thus, although the model of Ruderman (1991) can
%reproduce the observed radio properties of MSPs, it is inconsistent
%with the observed thermal X-ray pulse profiles of MSPs. 

\begin{figure}[t!]
  \includegraphics[height=.35\textheight]{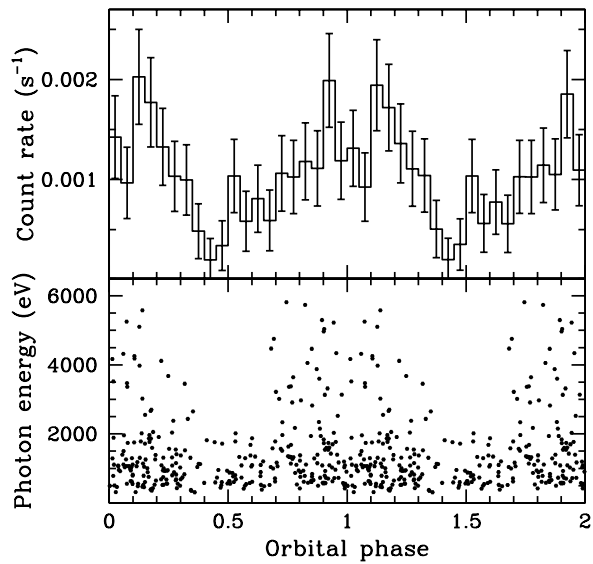} \caption{X-ray
  lightcurve of 47 Tuc W folded at the binary period ($P_b=3.2$
  h). Note the remarkable absence of hard X-ray photons between phases
  0.4 and 0.7, which can be explained by a geometric occultation of
  the intrabinary shock by the unusual main-sequence secondary star
  \citep[see][for details]{Bog05}.}
\end{figure}

\section{``Black Widow'' and ``Exchanged'' MSP Binaries}

A number of radio eclipsing binary MSPs, such as the canonical ``black
widow'' pulsar B1957+20 \citep{Stap03,Huang07}, 47 Tuc J and O
\citep{Bog06a}, and M71A \citep{Elsner07}, exhibit unresolved hard
non-thermal X-ray emission. This is likely synchrotron radiation from
a shock formed by the action of the pulsar wind on the companion star
\citep{Arons93}. X-ray studies (complemented by optical observations)
of these systems offer a useful diagnostic of the properties of MSP
winds.

The eclipsing binary MSPs PSR J1740--5340 in the globular cluster NGC
6397 \citep{Damico01,Grind02} and PSR J0024--7204W in 47 Tuc
\citep{Edmonds03,Bog05} exhibit very peculiar radio, optical, and
X-ray properties that set them apart from other MSPs (see Fig. 4).
The unusual behavior of these systems most probably stems from the
fact that they are bound to unevolved (main-sequence--like)
companions. This is at odds with the standard formation scenario of
MSPs, which requires the companion to evolve off the main sequence in
order for accretion and spin-up to commence \citep{Bhatt91}. It is
likely that the current binary partners were not the stars that spun
up the MSPs. As the dense environment of a globular cluster core is
conducive to close dynamical encounters between stars, it is very
likely that the present companion was acquired in a close
binary-binary exchange encounter. These ``exchanged'' binaries offer a
unique opportunity to study NSs transitioning from accretion to
rotation power.  In particular, the observed X-ray and optical
properties of the radio MSP PSR J0024--7204W are remarkably similar to
those of the low-mass X-ray binary and X-ray millisecond pulsar SAX
J1808.4--3658 in quiescence. This supports the conjecture that the
non-thermal X-ray emission and optical modulations seen in the SAX
J1808.4--3658 system in a quiescent state are due to interaction
between the wind from a reactivated rotation-powered pulsar and matter
from the companion star. The striking similarities between the two
systems provide further support for the long-sought connection between
millisecond radio pulsars and accreting neutron star systems.

%In particular if the
%it implies that SAX J1808.4--3658 is indeed on its way to becoming a
%rotation-powered MSP.

%We also examined the X-ray
%properties of the peculiar binary PSR J1740--5340 in the globular
%cluster NGC 6397. The hard non-thermal X-ray emission from this system
%is likely produced by the interaction of the pulsar wind with gas
%outflow from the secondary star.

\section{Non-thermal X-ray Pulsations}

A very small subset of MSPs show remarkably sharp X-ray pulsations,
indicative of non-thermal magnetospheric emission processes.  These
MSP include the exceptionally energetic PSRs B1937+21 \citep{Beck99},
B1821--24 \citep{Rut04}, and J0218+4232 ($\dot{E}=2.4\times10^{35}$
ergs s$^{-1}$, $B_{\rm surf}=4.3\times10^8$ G) \citep{Webb04} as well
as the mildly recycled relativistic binary MSP PSR J0737--3039A
\citep{Chat07}.  This striking difference in X-ray spectral properties
compared to most other MSPs implies a substantially different physical
regime in the pulsar magnetosphere. Models of pulsars suggest that the
combination of spin properties of these four pulsars place them near
the locus of theoretical curvature radiation (CR) pair-production
death lines \citep{Hard02b} on the $P-\dot{P}$ diagram.  Therefore, in
these of pulsars, pair-production in the magnetosphere probably occurs
via curvature radiation emitted by accelerating electrons. On the
other hand, most MSPs (such as those in 47 Tuc), lie well below the
theoretical CR pair production death lines. In these objects, inverse
Compton scattering (ICS) is the primary e$^{\mp}$ production mechanism
and the conditions in the magnetosphere favor substantial heating of
the polar caps by a return flow of e$^{+}$ \citep{Hard02}.

\section{Future Prospects}

%A deep Cycle 9 \textit{Chandra} ACIS-S observation of the 11 MSPs in
%the globular cluster M28.

Due to the inherent faintness of MSPs, at present, only a handful of
nearby objects can be studied in great detail \citep{Zavlin06}. The
next generation of X-ray facilities, \textit{Constellation-X} and
\textit{XEUS}, are expected to have $\sim$100 times greater
sensitivity than which would enable studies of a much greater sample
of MSPs. More importantly, these telescopes could be employed in blind
timing searches for nearby ($<$2 kpc) MSPs that are not observable at
radio frequencies due to unfavorable viewing geometry.  This is
possible because the effect of light bending combined with the
(nearly) antipodal configuration of the two MSP hot spots ensure that
the thermal radiation is seen by all distant observers for any
combination of $\alpha$ and $\zeta$.  In contrast, at radio
frequencies a pulsar is not observable if $|\alpha-\zeta|$ exceeds the
opening half-angle $\rho$ of the radio emission cone.
%  With the
%current generation of X-ray observatories (\textit{Chandra} and
%\textit{XMM-Newton}) this endeavor is difficult due to the intrinsic
%faintness of MSPs \citep{Cam07} and the relatively low X-ray pulsed
%fractions of these sources ($\le$50\%). On the other hand, with
%\textit{Constellation-X} and \textit{XEUS}, the great increase in
%sensitivity makes such a survey feasible.
%This brings forth the intriguing prospect of detecting and identifying
%such radio quiet MSPs in X-rays using pulsation searches.

Figure 5 shows the X-ray pulsed fraction of a 10 km, 1.4 M$_{\odot}$
MSP as a function of $\alpha$ and $\zeta$. Also shown are lines
delineating the region of the $\alpha-\zeta$ plane for which a pulsar
with a given radio emission cone width is observable at radio
frequencies.  If we assume a uniform distribution of pulsar
obliquities ($\alpha$) and viewing angles ($\zeta$), for $\rho\simeq
30^{\circ}$ a substantial portion ($\sim$45\%) of the MSP population
is invisible to us in the radio.  On the other hand, if we consider an
X-ray timing survey with a limiting pulsed fraction sensitivity of
$\sim$10\%, only $\sim$5--15\% (depending on the mass and radius of
the star) of the MSPs will go undetected as pulsed sources though they
will still be detected as X-ray sources \citep{Bog07b}.  The Galactic
population of MSPs may in fact be preferentially clustered in a
certain range of $\alpha$ due to the poorly understood effects of the
accretion and magnetic field reduction processes during the LMXB phase
on the NS.  A deep X-ray timing survey of nearby ($<$1-2 kpc) MSPs
may, in principle, reveal whether this is indeed the case.

\begin{figure}
  \includegraphics[angle=270,width=.35\textheight]{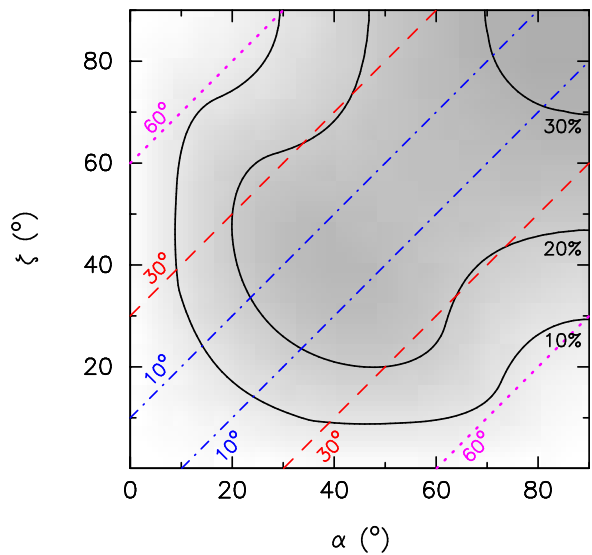}
  \caption{The $\alpha-\zeta$ plane for MSPs. The solid lines
  represent contours of constant thermal X-ray pulsed fraction for
  $R=10$ km and $M=1.4$ M$_{\odot}$. The dot-dashed, dashed, and
  dotted lines show 10$^{\circ}$, 30$^{\circ}$, and 60$^{\circ}$,
  respectively, opening half-angles of the radio emission cone.}
\end{figure}

\section{Conclusions}

To date, nearly 40 MSPs have been detected at X-ray energies, 19 of
which are found in the globular cluster 47 Tuc \citep{Bog06a}. It has
become apparent that X-ray spectroscopic and timing observations of
MSPs represent a very promising approach towards elucidating important
pulsar properties that are not measurable by other observational means
such as radio timing observations. Further study of these sources will
provide more important information regarding the X-ray properties of
the Galactic population of MSPs, as well as valuable insight into the
fundamental properties of NSs.

%%%%%%%%%%%%%%%%%%%%%%%%%%%%%%%%%%%%%%%%%%%%%%%%
%% BACKMATTER
%%%%%%%%%%%%%%%%%%%%%%%%%%%%%%%%%%%%%%%%%%%%%%%%

\begin{theacknowledgments}

\end{theacknowledgments}

We would like to thank G. Rybicki, C. Heinke, M. van den
Berg, F. Camilo, P. Freire, A. Harding, and W. Becker
for their valuable contribution to the work described herein. We also
extend our thanks to the conference organizers for their hospitality.

\end{document}